\documentclass[preprintnumbers,amssymb,nofootinbib]{revtex4}

\usepackage{graphicx}
\usepackage{epsfig}
\usepackage{dcolumn}
\usepackage{bm}
\usepackage{amsmath}
\usepackage{amssymb}

\def\beq{\begin{equation}}
\def\eeq{\end{equation}}
\def\bea{\begin{eqnarray}}
\def\eea{\end{eqnarray}}
\def\beq{\begin{equation}}
\def\eeq{\end{equation}}
\def\be{\begin{equation}}
\def\ee{\end{equation}}

\def\gev{{\rm GeV}}

\begin{document}  

\title{$b \to s$ decays in a model with $Z$-mediated flavor changing neutral current}

\author{Ashutosh Kumar Alok}
\email{akalok@iitj.ac.in}
\affiliation{Indian Institute of Technology Rajasthan, Jodhpur 342011, India}

\author{Shireen Gangal}
\email{shireen.gangal@desy.de }
\affiliation{Theory Group, Deutsches Elektronen-Synchrotron (DESY), D-22607 Hamburg, Germany\vspace{2ex}}

\date{\today} 
\preprint{DESY 12-153}

\begin{abstract}
In the scenario with $Z$ mediated flavor changing neutral current occurring at the tree level due to the addition of a vector-like
isosinglet down-type quark $d'$ to the SM particle spectrum, we perform a $\chi^2$ fit using the flavor physics data and 
obtain the best fit value along with errors of the tree level $Z{\bar b}s$ coupling, $U_{sb}$.
The fit indicates that the new physics coupling is constrained to be small:  
we obtain $|U_{sb}|$ $\le$ $3.40 \times 10^{-4}$ at 3$\sigma$. Still this does allow for the possibility 
of new physics signals in some of the observables such as semileptonic CP asymmetry 
in $B_s$ decays.
\end{abstract}

\maketitle 

\newpage

\section{Introduction}
The Standard Model (SM) of the electroweak interactions successfully 
explains most of the experimental data to date. However in recent years, 
there have been quite a few measurements of
quantities in $B$ decays which differ from the predictions of the
SM. For example, in $B \to \pi K$, the
SM has some difficulty in accounting for all the experimental
measurements \cite{piKupdate}.  The measured indirect (mixing-induced)
CP asymmetry in some $b \to  s$ penguin decays is found not to be
identical to that in ${\bar B}\to J/\psi K_{\scriptscriptstyle S}$ \cite{btos-1,btos-2,btos-3},
counter to the expectations of the SM. The measurement of 
indirect CP asymmetry in ${\bar B}_s\to J/\psi \phi$  
by the CDF and D\O\ collaborations shows a deviation from the SM prediction \cite{Aaltonen:2007he,Aaltonen:2007gf,hfag}
\footnote{The recent LHCb update does not confirm this result \cite{LHCbupdate1}.
Their measurement is consistent with the SM prediction}.
The observation of the anomalous dimuon
charge asymmetry by the D\O\ collaboration \cite{D0-dimuon,Abazov:2010hj,Abazov:2011yk} also
points towards some new physics in $B_s$ mixing that affects the
lifetime difference and mixing phase involved therein \cite{Dighe:2007gt,Dighe:2010nj}.  
A further hint of new physics has been seen 
in the exclusive semileptonic decay ${\bar B}\to {\bar{K}^*} \mu^+ \mu^-$:
the forward-backward asymmetry ($A_{FB}$) has been found to deviate somewhat from the predictions of
the SM \cite{Belle-newKstar,BaBar-Kstarmumu,Aaltonen:2011ja,Alok:2009tz}
\footnote{The recent LHCb update does not confirm this result \cite{LHCbupdate2}.
Their measurement of the $A_{FB}$ distribution is
consistent with the SM prediction, except in the high-$q^2$ region.}.
Though the disagreements are only at the level of $\sim$ 2-3$\sigma$, and hence not statistically significant, 
they are intriguing since they all appear in $b \to s$ transitions.
Therefore the study of new physics effects in various $b \to s$ observables is crucially important.

A minimal extension of SM can be obtained by adding a vector-like
isosinglet up-type or down-type quark to the SM particle spectrum 
\cite{delAguila:1985mk,Branco:1986my,delAguila:1985ne,Nir:1990yq,Branco:1992wr,
Barger:1995dd,Silverman:1995rk,Lavoura:1992qd,Silverman:1991fi,Barenboim:1997pf,
Barenboim:2000zz,Barenboim:2001,Chen:2010aq,Alok:2010ij,Botella:2012ju,Okada:2012gy}. Such exotic
fermions can appear in $E_6$ grand unified theories as well in models with 
large extra dimensions. Here we consider the extension of SM by adding a vector like down-type 
quark $d'$. The ordinary $Q_{em} =-1/3$ quarks mix with the $d'$.  Because the $d'_L$ has a different
$I_{3L}$ from $d_L$, $s_L$ and $b_L$, $Z$-mediated FCNC's (ZFCNC) appear at tree level in
the left-handed sector.  In particular, a $Z{\bar b}s$ coupling can be
generated:
\be
{\cal L}^{ Z}_{ FCNC} = -\frac{g}{2 \cos\theta_{
W}} U_{sb} \, \bar s \gamma^\mu P_L b \, Z_\mu
+ {\rm h.c.} 
\label{Usb} 
\ee
This coupling leads to a new physics contribution to $b\to s$ transition 
(such as $B_s$-$\bar{B_s}$ mixing, $b\to s \mu^+ \,\mu^-$ \& 
$b \to s \nu \bar{\nu}$ decays, etc) at the tree level. This tree level coupling
$U_{sb}$ can be constrained by various measurements in the $b \to s$ sector. 

In this paper we consider observables such as $B_s$-$\bar{B_s}$ mixing, 
branching ratios of $\bar{B} \to X_s \mu^+ \mu^-$, $\bar{B_s} \to \mu^+ \mu^-$ 
and $\bar{B}  \to X_s \nu \bar{\nu}$ to constrain the new physics coupling $U_{sb}$. 
Instead of obtaining the usual scatter plot which shows the allowed ranges of the $U_{sb}$ parameter space, 
we perform a $\chi^2$ fit which provides us the best fit value of $U_{sb}$ along with the errors. We then study 
the effect of tree level $Z{\bar b}s$ coupling on the indirect $CP$ asymmetry in $B_s \to \psi \phi$, anomalous dimuon
charge asymmetry $a^s_{sl}$, forward-backward (FB) asymmetry in $\bar{B} \to X_s \mu^+ \mu^-$ and
the branching ratio of $\bar{B_s} \to \tau^+ \tau^-$. We show that the various measurements in the $b \to s$ sector put strong constraint on the allowed values of $U_{sb}$. However it is still possible to have new physics signals in some $b \to s$ observables. 

The paper is organized as follows. In Sec. \ref{sec:method}, we discuss the methodology for the fit. 
In Sec. \ref{sec:results}, we present the results of the fit. 
In Sec. \ref{sec:pred}, we obtain predictions for various $b \to s$ 
observables. Finally in Sec. \ref{sec:concl}, we present our conclusions.  

\section{Method}
\label{sec:method}

As $U_{sb}$ denotes the $Z{\bar b}s$ coupling
generated in the $Z$FCNC model, the parameters of
the model are therefore the magnitude and the phase of this coupling,
$|U_{sb}|$ and $\phi_{sb} \equiv \arg U_{sb}$.

In order to obtain constraints on the new physics coupling $U_{sb}$, we perform
a $\chi^2$ fit using the CERN minimization code MINUIT \cite{minuit}.  
The fit includes observables that have relatively small hadronic uncertainties:
(i) the branching ratio of $\bar{B} \to X_s \mu^+ \mu^-$ in the low- and 
high-$q^2$ regions,
(ii) the branching ratio of $\bar{B_s} \to \mu^+ \mu^-$,
(iii) the ratio of the branching ratio of $\bar{B_s}  \to \mu^+ \mu^-$ 
and the mass difference in $B_s$ system,
(iv) the branching ratio of $\bar{B}  \to X_s \nu \bar{\nu}$.
We include both experimental errors and theoretical uncertainties in the fit. 
In the following subsections, we discuss various observables used as a constraint.

\subsection { $\bar{B} \to X_s\,  \mu^+ \,\mu^-$}

The effective Hamiltonian for the quark-level transition
$b \to s\,  \mu^+ \,\mu^-$ in the SM can be written as
\be
{\cal H}_{eff} =  - \frac{4 G_F}{\sqrt{2}} V^*_{ts}V_{tb}
\sum_{i=1}^{10} C_i(\mu) \,  O_i(\mu)\;,
\ee
where the form of the operators $O_i$ and the expressions for
calculating the coefficients $C_i$ are given in
Ref.~\cite{Buras:1994dj}. The operator $O_i$, $i=1,6$ can contribute indirectly to 
$b \to s\,  \mu^+ \,\mu^-$ and their effects are included in the 
effective Wilson coefficients $C_9$ and $C_7$ \cite{Buras:1994dj,Altmannshofer:2008dz}.

The $Z{\bar b}s$ coupling generated in the $Z$FCNC model changes the values of the 
Wilson coefficients $C_{9,10}$.  The Wilson coefficients $C^{\rm tot}_{9,10}$ in
the $Z$FCNC model can be written as
\bea
C^{\rm tot}_{9} &=& C_9^{\rm eff} - \frac{\pi}{\alpha}\frac{U_{sb}}{V^*_{ts}V_{tb}} (4\sin^2 \theta_{W}-1)\,\\
C^{\rm tot}_{10} &=& C_{10}-\frac{\pi}{\alpha}\frac{U_{sb}}{V^*_{ts}V_{tb}}\,.
\eea
Here $V^*_{ts}V_{tb}\simeq -0.0403\, e^{-i\, 1^\circ}$. We use the SM Wilson coefficients as given in
Ref.~\cite{Altmannshofer:2008dz}.

The calculation of branching ratio gives
\be
{\cal{BR}}(\bar{B}  \to X_s\, \mu^+\, \mu^-)
 = \frac{\alpha^2 {\cal{BR}}(B\rightarrow X_c e {\bar \nu})}
 {4 \pi^2 f(\hat{m_c})\kappa(\hat{m}_c)} 
 \frac{|V^*_{ts}V_{tb}|^2}{|V_{cb}|^2} \int D(z) dz \;,
\label{eq:brincl}
\ee
where
\be
D(z) = (1-z)^2 \Big[ (1+2z)\left( |C_9^{\rm tot}|^2 + |C_{10}^{\rm tot}|^2 \right)
      + 4 \left(1+\frac{2}{z}\right) |C_7^{\rm eff}|^2
     +12 {\rm Re}(C_7^{\rm eff} C_{9}^{\rm tot*})\Big]\;.
\label{eq:dz}
\ee
Here $z \equiv q^2/m_{b}^{2}\equiv (p_{\mu^+}+p_{\mu^-})^2/m_{b}^{2} $ and $\hat{m}_q=m_q/m_b$ for all quarks $q$.
The expressions for the phase-space factor $f(\hat{m_c})$ and the
$1$-loop QCD correction factor $\kappa(\hat{m_c})$ are given
in \cite{Nir:1989rm}.

The theoretical prediction for the branching ratio of $\bar{B}  \to X_s\, \mu^+\, \mu^-$ 
in the intermediate $q^2$ region ($7$~GeV$^2 \le q^2 \le
12$~GeV$^2$) is rather uncertain due to the nearby charmed
resonances. The predictions are relatively cleaner in the low-$q^2$
($1 \,{\rm GeV^2} \le q^2 \le 6\, {\rm GeV^2}$) and the high-$q^2$
($14.4\, {\rm GeV^2} \le q^2 \le m_b^2$) regions. We therefore consider
both low-$q^2$ high-$q^2$  region in the fit.

We define $\chi^2$ as
\bea
\chi^2_{\bar{B} \to X_s\, \mu^+\, \mu^-: \rm low} &=& \Big( \frac{D_{\rm low}-5.69947}{1.82522} \Big)^2\;,\\
\chi^2_{\bar{B} \to X_s\, \mu^+\, \mu^-: \rm high} &=& \Big( \frac{D_{\rm high}-1.56735}{0.635465} \Big)^2\;,
\eea
where 
\bea
D_{\rm low} =\int_{\frac{1}{m_b^2}}^{\frac{6}{m_b^2}} D(z) dz 
&=& {\cal{BR}}(\bar{B} \to X_s  \mu^+ \,\mu^-)_{\rm low} \frac{ 4 \pi^2\, f(\hat{m_c})\,\kappa(\hat{m}_c)}{\alpha^2 {\cal{BR}}(B\rightarrow X_c e {\bar \nu})} 
\frac{|V_{cb}|^2}{|V^*_{ts}V_{tb}|^2}=5.69947 \pm 1.82522\;,\\
D_{\rm high} =\int_{\frac{14.4}{m_b^2}}^{(1-\frac{m_s}{m_b})^2} D(z) dz 
&=& {\cal{BR}}(\bar{B} \to X_s  \mu^+ \,\mu^-)_{\rm high} \frac{ 4 \pi^2\, f(\hat{m_c})\,\kappa(\hat{m}_c)}{\alpha^2 {\cal{BR}}(B\rightarrow X_c e {\bar \nu})} 
\frac{|V_{cb}|^2}{|V^*_{ts}V_{tb}|^2}=1.56735 \pm 0.635465\;.
\label{eq:brinclh}
\eea
Here we have added an overall corrections of 30\% to the theoretical prediction of ${\cal{BR}}(B \to X_s  \mu^+ \,\mu^-)_{\rm high}$, which includes the non-perturbative corrections.

\begin{table}[htbp]
\begin{center}
\begin{tabular}{|c|c|}
\hline
\hline
$\eta_B = 0.5765\pm 0.0065$ \cite{buras1} & ${\cal{BR}}(\bar{B_s}\to \mu^+\, \mu^-) = (0.0\pm 2.30)\times 10^{-9}$ \cite{LHCbupdate}\\
  $f_{bs} = 0.229 \pm 0.006\, \gev$  \cite{CKMfitter,Lenz:2012az} & ${\cal{BR}}(\bar{B}\to X_s\, \mu^+\, \mu^-)_{\rm low} = (1.60 \pm 0.50)\times 10^{-6}$ \cite{Aubert:2004it,Iwasaki:2005sy} \\
$B_{bs}= 1.291 \pm 0.043$ \cite{CKMfitter,Lenz:2012az} & ${\cal{BR}}(\bar{B}\to X_s\, \mu^+\, \mu^-)_{\rm high} = (0.44 \pm 0.12)\times 10^{-6}$ \cite{Aubert:2004it,Iwasaki:2005sy}\\
$\Delta{M_s} = (17.69\pm 0.08)\, ps^{-1}$  \cite{Asner:2010qj} & ${\cal{BR}}(\bar{B}\to X_s \nu \nu) = (0.0 \pm 40)\times 10^{-5}$ \cite{Barate:2000rc}\\
$\frac{|V^*_{ts}V_{tb}|}{|V_{cb}|} = 0.967\pm0.009 $ \cite{CKMfitter} & $m_t(m_t) = 163.5\, \gev$ \\
$|V^*_{ts}V_{tb}|=-(0.0403\pm0.0009)$ & $m_c/m_b=0.29 \pm 0.02$\\
${\cal{BR}}(B\to X_c \ell \nu) = (10.61 \pm 0.17)\times 10^{-2}$  & $\tau_{B_s}=(1.520 \pm 0.020)\, ps^{-1}$ \cite{LHCbupdate1} \\
\hline
\hline
\end{tabular}
\caption{Inputs that we use in order to constrain $|U_{sb}|$-$\phi_{sb}$ parameter space, when not 
explicitly stated, we take the inputs from Particle Data Group \cite{pdg}.}
\label{tab1}
\end{center}
\end{table}

\subsection{$\bar{B_s} \to \mu^+ \mu^-$}

The purely leptonic decay $\bar{B_s} \to \mu^+ \mu^-$ is 
chirally suppressed within the SM. The SM prediction for 
the branching ratio is $(3.35 \pm 0.32)\times 10^{-9}$ \cite{Blanke:2006ig}. 
Recently LHCb collaboration reported a very strong upper bound on 
the branching ratio of $\bar{B_s} \to \mu^+ \mu^-$, 
which is $3.8 \times 10^{-9}$ at 90\% C.L. \cite{LHCbupdate}.

The branching ratio of $\bar{B_s} \to \mu^+ \mu^-$ in the $Z$FCNC model is given
by
\beq
{\cal{BR}}({\bar B}_s \to \mu^+ \,\mu^-) = \frac{G^2_F \alpha^2 M_{B_s} m_\mu^2 f_{bs}^2 \tau_{B_s}}{16 \pi^3}
|V_{ts}^* V_{tb}|^2 \sqrt{1 - \frac{4 m_\mu^2}{M_{B_s}^2}} |C_{10}^{\rm tot}|^2\;.
\label{bmumu-BR}
\eeq

We define $\chi^2$ as
\beq
\chi^2_{\bar{B_s} \to \mu^+ \mu^-} = \Big( \frac{|C_{10}^{\rm tot}|^2 - 0.0}{13.5408} \Big)^2\;,
\eeq
with 
\beq
|C_{10}^{\rm tot}|^2 = \frac{16 \pi^3 {\cal{BR}}({\bar B}_s \to \mu^+ \,\mu^-)}{G^2_F \alpha^2 M_{B_s} m_\mu^2 f_{bs}^2 \tau_{B_s}
|V_{ts}^* V_{tb}|^2 \sqrt{1 - \frac{4 m_\mu^2}{M_{B_s}^2}}} =0.0 \pm 13.5408 \;.
\eeq

\subsection {Ratio of ${\cal{BR}}(\bar{B_s}  \to \mu^+ \mu^-)$ 
and the mass difference in the $B_s$ system}
The mass difference $\Delta M_s$ is given by
\beq
\Delta M_s=2|M_{12}^{{\rm SM}} |\;.
\eeq
The SM contribution to $M_{12}^s$ is
\bea
M_{12}^{s,{\rm SM}} = \frac{G_F^2}{12 \pi^2}  (V^*_{ts}V_{tb})^2 M_W^2 M_{B_s}
 \eta_ B f_{B_s}^2 B_{B_s}  E(x_t) ~,
\eea
where $x_t = m_t^2/M_W^2$ and $\eta_B$ is the QCD correction.  
The loop function $E(x_t)$ is given by
\bea
  E(x_t) = \frac{-4 x_t + 11 x_t^2- x_t^3  }{4(1-x_t)^2}
+\frac{3 x_t^3 \ln x_t}{2(1-x_t)^3} ~.
\eea

The mass difference $\Delta M_s$ in the $Z$FCNC model is given
by~\cite{Barenboim:1997pf}
\bea
 \Delta M_s = \frac{G^2_F}{6\pi^2} |V^*_{ts}V_{tb}|^2 M_W^2 M_{B_s} \eta_ B  f_{bs}^2 B_{bs} |E(x_t)| |\Delta_s|\;.
\label{DMsZFCNC}
\eea
$\Delta_s$ is given by
\bea
 \Delta_s= 1 + a \left(\frac{U_{sb}}{V^*_{ts}V_{tb}} \right) - b \left(\frac{U_{sb}}{V^*_{ts}V_{tb}} \right)^2 \;,
\label{deltas}
\eea
where
\bea
  a = 4 \frac{C(x_t)}{E(x_t)} ~~,~~~~
  b = \frac{2 \sqrt{2} \pi^2}{G_F M_W^2 E(x_t)} ~.
\eea
The loop function $C(x_t)$ is given by~\cite{Barenboim:1997pf}
\beq
  C(x_t)  =  \frac{x_t}{4} \left[\frac{4-x_t}{1-x_t} 
+ \frac{3 x_t \ln x_t  }{(1-x_t)^2} \right] ~.
\eeq
The term in Eq.~(\ref{DMsZFCNC}) proportional to $a$ is obtained from
a diagram with both SM and new physics $Z$ vertices; that proportional to $b$
corresponds to the diagram with two new physics $Z$ vertices.

Dividing Eq.~(\ref{bmumu-BR}) by Eq.~(\ref{DMsZFCNC}), we get
\beq
\frac{{\cal{BR}}({\bar B}_s \to \mu^+ \,\mu^-) }{ \Delta M_s}= \frac{3\alpha^2 \tau_{B_s} m_\mu^2 }{8 \pi M_W^2 \eta_ B B_{bs} |E(x_t)| } 
\sqrt{1 - \frac{4 m_\mu^2}{M_{B_s}^2}} \frac{|C_{10}^{\rm tot}|^2}{|\Delta_s|}
\label{ratio:br-mix}
\eeq

We define $\chi^2$ as
\beq
\chi^2_{BR-mix} = \Bigg( \frac{\frac{|C_{10}^{\rm tot}|^2}{|\Delta_s|} - 0.0}{13.6328} \Bigg)^2\;,
\eeq
with
\beq
 \frac{|C_{10}^{\rm tot}|^2}{|\Delta_s|} = \frac{{\cal{BR}}({\bar B}_s \to \mu^+ \,\mu^-) }{ \Delta M_s \sqrt{1 - \frac{4 m_\mu^2}{M_{B_s}^2}}} \frac{8 \pi M_W^2 \eta_ B B_{bs} |E(x_t)|} {3\alpha^2 \tau_{B_s} m_\mu^2 }=0.0\pm 13.6328\;.
\eeq

\subsection{$\bar{B} \to X_s \nu \bar{\nu}$}
The effective Hamiltonian for the decay $\bar{B} \to X_s \nu \bar{\nu}$ is given by
\beq
H_{eff}= \frac{G_F}{\sqrt{2}} \frac{\alpha}{2 \pi \sin^2 \theta_W}
V^*_{ts}V_{tb}X_0(x_t) (\bar{s} b)_{V-A} (\bar{\nu} \nu)_{V-A} + {\rm h.c.}\;,
\eeq
with 
\beq
X_0(x_t) = \frac{x_t}{8}\Big[\frac{2 + x_t}{x_t -1}+\frac{3x_t-6}{(x_t-1)^2}\ln x_t\Big]\;.
\eeq
The presence of tree level $Z{\bar b}s$ coupling changes the value of the 
structure function $X_0(x_t)$.  The  structure function within the $Z$FCNC model 
can be written as
 \beq
X'_0(x_t) = X_0(x_t) + \Big(\frac{\pi \sin^2 \theta_W}{\alpha V^*_{ts}V_{tb}}\Big) U_{sb}\;.
\eeq
The branching ratio of $\bar{B} \to X_s \nu \bar{\nu}$ is given
by \cite{Grossman:1995gt,Buras:1997fb}
\beq
{\cal{BR}}(\bar{B}  \to X_s \nu \bar{\nu}) ={\cal{BR}}(B\rightarrow X_c e {\bar \nu}) \frac{\tilde{C}^2 \bar{\eta}}{|V_{cb}|^2  f(\hat{m_c})\kappa(\hat{m}_c)} \;,
\label{eq:brinclnunu}
\eeq
where $\tilde{C}^2$ is given by
\beq
\tilde{C}^2 =\frac{\alpha^2}{2\pi^2 \sin^4 \theta_W} |V^*_{ts}V_{tb} X'_0(x_t)|^2.
\eeq

We define $\chi^2$ as
\beq
\chi^2_{\bar{B}  \to X_s \nu \bar{\nu}} = \Bigg(\frac{|V^*_{ts}V_{tb} X'_0(x_t)|^2 - 0.0}{0.069157} \Bigg)^2\;,
\eeq
with
\beq
|V^*_{ts}V_{tb} X'_0(x_t)|^2 = \frac{{\cal{BR}}(\bar{B}  \to X_s \nu \bar{\nu})}{{\cal{BR}}(B\rightarrow X_c e {\bar \nu})} \frac{2\pi^2 \sin^4 \theta_W |V_{cb}|^2  f(\hat{m_c})\kappa(\hat{m}_c)}{\bar{\eta} \alpha^2}=0.0 \pm 0.069157\;.
\eeq
Here we have used the present upper bound ${\cal{BR}}(\bar{B} \to X_s \nu \bar{\nu})< 64 \times 10^{-5}$ at 90\% C.L. \cite{Barate:2000rc} which can be written as $(0.0 \pm 40)\times 10^{-5}$. 

Therefore the total $\chi^2$ can be written as
\bea
\chi^2_{\rm total} = 
\chi^2_{\bar{B}  \to X_s\, \mu^+ \,\mu^-:\rm low} +
 \chi^2_{\bar{B}  \to X_s\, \mu^+ \,\mu^-:\rm high} + \chi^2_{\bar{B_s} \to \mu^+ \mu^-}  + \chi^2_{BR-mix}  + \chi^2_{\bar{B}  \to X_s \nu \bar{\nu}}\;.
\eea

\section{Results of the Fit}
\label{sec:results}


\begin{table}
\begin{tabular}{ccc}
\hline
\hline
Parameter  & Value\\
\hline
$|U_{sb}|$  & $(0.90 \pm 0.83) \times 10^{-4}$ \\
$\phi_{sb}$ & $(0.00\pm 181.34)^\circ$ \\
\hline
$\chi^2/d.o.f.$ & $1.72/3$ \\
\hline
\hline
\end{tabular}
\caption{The results of the fit to the parameters of $Z$FCNC model.
\label{tab1:parameters}}
\end{table}
The results of these fits are presented in Table \ref{tab1:parameters}. It may be observed that 
the $\chi^2$ per degree of freedom is small, indicating that the fit is good. We observe that the present flavor data put strong constraint on $Z\bar{b}s$ coupling. 
At 3$\sigma$, we obtain $|U_{sb}|\leq 3.40 \times 10^{-4}$.

\section{Predictions}
\label{sec:pred}

\begin{table}
\begin{tabular}{|c|c|c|}
\hline
& \multicolumn{1}{c}{} Predictions &    \\
\cline{2-3} Observables & SM  &  ZFCNC    \\
\hline
$\phi^{\Delta}_s $ (rad)& $0$ & $  (0.00\pm 0.03)$   \\
\hline
$|\Delta_s|$ & $1$ & $ 1.01 \pm 0.01$ \\
\hline
$a^s_{sl}\times 10^5$ & $(1.92\pm 0.67) $ & $ (1.98\pm 13.88)$  \\
\hline
$Br(B_s \to \tau^+ \tau^+)\times 10^{7}$ & $5.74\pm 0.27$ & $3.34\pm 1.92$\\
\hline
$(q^2)_0^{\rm incl}\, {\rm GeV^2}$ & $3.33 \pm 0.25$ & $3.38 \pm 0.26$ \\
\hline
\end{tabular}
\caption{ZFCNC predictions for potential observables.}
\label{predictions}
\end{table}

\subsection{Semileptonic asymmetry $a^s_{sl}$}
The expression for the semileptonic asymmetry $a^s_{sl}$ is given by
\be
a^s_{sl} = \frac{|\Gamma^s_{12}|}{|M^s_{12}|} \sin\phi_s = \frac{|\Gamma^s_{12}|}{|M^{s,SM}_{12}|} \frac{\sin\phi_s}{|\Delta_s|}\,,
\label{asls}
\ee
where the CP violating phase $\phi_{s}$ is defined by the following equation, 
\be
\phi_s \equiv Arg\left[- \frac{M^s_{12}}{\Gamma^s_{12}}\right]\, .
\ee
The parameter $\Delta_s$ takes into account the new physics effects in mixing and is defined as
\be
M^s_{12} = M^{s,SM}_{12}(1 + \frac{M^{s,NP}_{12}}{M^{s,SM}_{12}}) = M^{s,SM}_{12} \Delta_s= M^{s,SM}_{12} |\Delta_s| e^{\phi^{\Delta}_s}.
\label{delm}
\ee
Thus $\phi_{s}$ can be written as  
\be
\phi_{s} = \phi^{\Delta}_s +  \phi_{s}^{\rm SM}\;,
\label{eq:phis}
\ee
where $\phi_{s}^{\rm SM}=(3.84 \pm 1.05) \times 10^{-3}$ \cite{Lenz:2006hd}. Also, one has \cite{uli_lenz,Lenz:2012mb}
\be
\frac{|\Gamma^s_{12}|}{|M^{s,SM}_{12}|} = (5.0 \pm 1.1)\times 10^{-3}.
\label{rat:gm12}
\ee
The predictions for $\phi^{\Delta}_s $, $|\Delta_s|$  and $a^s_{sl} $ in ZFCNC model are given in Table ~\ref{predictions}. We see that it is possible to have large deviations in $\phi_s$ (and hence $a^s_{sl} $) from its SM predictions.

\subsection{Zero of Forward-Backward asymmetry}

The FB asymmetry of muons in $\bar{B}  \to X_s\, \mu^+\, \mu^-$ is obtained by
integrating the double differential branching ratio
($\frac{d^2 {\cal{BR}}}{dz\, d\cos\theta}$) with respect to the angular variable $\cos\theta$ \cite{Ali:1991is}
\bea
A_{FB}(z)= \frac{\int_0^{1}d\,cos\theta \frac{d^2{\cal{BR}}}{dz \
d\cos\theta}-\int_{-1}^{0}d\,\cos\theta \frac{d^2{\cal{BR}}}{dz\ d\cos\theta}}
{\int_0^{1}d\,cos\theta \frac{d^2{\cal{BR}}}{dz\ dcos\theta}+\int_{-1}^{0}dcos\theta \frac{d^2{\cal{BR}}}{dz
\ d\cos\theta}}\;,
\eea
where  $\theta$
is the angle between the momentum of the $\bar{B}$-meson
and that of $\mu^+$ in the dimuon center-of-mass frame.

Within the ZFCNC model, FB asymmetry in $\bar{B}  \to X_s\, \mu^+\, \mu^-$ is given by
\be
A_{FB}(z) =\frac {-3 E(z)}{D(z)}\;,
\ee
where $D(z)$ is given in Eq.~\ref{eq:dz} and $E(z)$ by
\be
E(z)= {\rm Re}(C_9^{\rm tot}C_{10}^{\rm tot*})z
  + 2{\rm Re}(C_7^{\rm eff}C_{10}^{\rm tot*})\;.
\ee

Zero of $A_{FB}(z)$ is determined by
\be 
E(z)= {\rm Re}(C_9^{\rm tot}C_{10}^{\rm tot*})z
  + 2{\rm Re}(C_7^{\rm eff}C_{10}^{\rm tot*})=0\;.
\ee

The prediction for $(q^2)_0^{\rm incl}$ in ZFCNC model is given in Table ~\ref{predictions}. One can see that large deviations from SM prediction is not possible.

\subsection{${\cal{BR}}(\bar{B_s} \to \tau^+ \, \tau^-)$}

The branching ratio of $\bar{B_s} \to \tau^+ \, \tau^-$ in the $Z$FCNC model is given
by
\be
{\cal{BR}}({\bar B}_s \to \tau^+ \, \tau^-)  = \frac{3\alpha^2 \tau_{B_s} m_\tau^2 }{8 \pi M_W^2 \eta_ B B_{bs} |E(x_t)| } 
\sqrt{1 - \frac{4 m_\tau^2}{M_{B_s}^2}} \frac{|C_{10}^{\rm tot}|^2}{|\Delta_s|} \Delta M_s
\label{bstautau-BR}
\ee

The prediction for ${\cal{BR}}(\bar{B_s} \to \tau^+ \, \tau^-)$ in ZFCNC model is given in Table ~\ref{predictions}. We see that it is possible to have large suppression in ${\cal{BR}}(\bar{B_s} \to \tau^+ \, \tau^-)$ as compared to its SM prediction.

\section{Conclusion}
\label{sec:concl}

In this paper, we consider a minimal extension of the SM by adding a vector-like
isosinglet down-type quark $d'$ to the SM particle spectrum.
As a consequence, $Z$-mediated FCNC's appear at tree level in
the left-handed sector.  In particular,  we are interested in $Z{\bar b}s$ coupling 
which leads to a new physics contribution to $b\to s$ transition 
such as $B_s$-$\bar{B_s}$ mixing, $b\to s \mu^+ \,\mu^-$,  
$b \to s \nu \bar{\nu}$ decays, etc at the tree level. Using 
inputs from several observables in flavor physics, we perform 
a $\chi^2$ fit to constrain the tree level $Z{\bar b}s$ coupling, $U_{sb}$.
The fit takes into account both the theoretical as well as the experimental
uncertainties. 

We conclude the following:

\begin{itemize}
\item $\chi^2$ per degree of freedom is small, indicating that the fit is good. This is
expected as the SM itself is in good agreement with the data.

\item The present data put strong constraint on the $Z\bar{b}s$ coupling. At 3$\sigma$, 
$|U_{sb}|\leq 3.40 \times 10^{-4}$.

\item Despite the strong constraint on the $Z\bar{b}s$ coupling, it is possible to have new physics signals in some $b \to s $ observables such as semileptonic $CP$ asymmetry in $B_s$ decays. 
 
\end{itemize}


\section*{Acknowledgments}

We thank Soumitra Nandi for helpful collaboration on several parts of this analysis. 
The work of S. G. is supported by DFG Emmy Noether grant TA 867/1-1.



\begin{thebibliography}{10}

\bibitem{piKupdate} In the latest update of the $\pi K$ puzzle, it was
  seen that, although NP was hinted at in $B \to \pi K$ decays, it could
  be argued that the SM can explain the data, see  S.~Baek, C.~-W.~Chiang, D.~London,
  Phys.\ Lett.\  {\bf B675}, 59-63 (2009).
  [arXiv:0903.3086 [hep-ph]].

\bibitem{btos-1}
  H.~-Y.~Cheng, C.~-K.~Chua, A.~Soni,
  Phys.\ Rev.\  {\bf D72}, 094003 (2005).
  [hep-ph/0506268].

\bibitem{btos-2}
  G.~Buchalla, G.~Hiller, Y.~Nir, G.~Raz,
  JHEP {\bf 0509}, 074 (2005).
  [hep-ph/0503151].


\bibitem{btos-3}
  E.~Lunghi, A.~Soni,
  JHEP {\bf 0908}, 051 (2009).
  [arXiv:0903.5059 [hep-ph]].


\bibitem{Aaltonen:2007he}
  T.~Aaltonen {\it et al.} [ CDF Collaboration ],
  Phys.\ Rev.\ Lett.\  {\bf 100}, 161802 (2008).
  [arXiv:0712.2397 [hep-ex]].

\bibitem{Aaltonen:2007gf}
  T.~Aaltonen {\it et al.} [ CDF Collaboration ],
  Phys.\ Rev.\ Lett.\  {\bf 100}, 121803 (2008).
  [arXiv:0712.2348 [hep-ex]];
 D\O\ Collaboration, Conference Note 5933-CONF, May 28, 2009, 
http://www-d0.fnal.gov/Run2Physics/WWW/results/prelim/B/B58/B58.pdf.

\bibitem{hfag}
  D.~Asner {\it et al.} [ Heavy Flavor Averaging Group Collaboration ],
  [arXiv:1010.1589 [hep-ex]].

\bibitem{LHCbupdate1}
"Combination of $\phi_s$ measurements from $B_{s}^{0}\to J/\psi\phi$ and $B_{s}^{0}\to J/\psi f_{0}(980)$", 
LHCb-CONF-2011-056 ; "Tagged time-dependent angular analysis of $B_{s}^{0}\to J/\psi\phi$  decays at LHCb",  LHCb-CONF-2012-002.

\bibitem{D0-dimuon}
  V.~M.~Abazov {\it et al.}  [D0 Collaboration],
  Phys.\ Rev.\  D {\bf 82}, 032001 (2010)
  [arXiv:1005.2757 [hep-ex]].

\bibitem{Abazov:2010hj}
  V.~M.~Abazov {\it et al.} [ D0 Collaboration ],
  Phys.\ Rev.\ Lett.\  {\bf 105}, 081801 (2010).
  [arXiv:1007.0395 [hep-ex]].


\bibitem{Abazov:2011yk}
  V.~M.~Abazov {\it et al.}  [D0 Collaboration],
  arXiv:1106.6308 [hep-ex].

\bibitem{Dighe:2007gt}
  A.~Dighe, A.~Kundu, S.~Nandi,
  Phys.\ Rev.\  {\bf D76}, 054005 (2007).
  [arXiv:0705.4547 [hep-ph]].


\bibitem{Dighe:2010nj}
  A.~Dighe, A.~Kundu, S.~Nandi,
  Phys.\ Rev.\  {\bf D82}, 031502 (2010).
  [arXiv:1005.4051 [hep-ph]].



\bibitem{Belle-newKstar}
  J.~T.~Wei {\it et al.}  [BELLE Collaboration],
  Phys.\ Rev.\ Lett.\  {\bf 103}, 171801 (2009)
  [arXiv:0904.0770 [hep-ex]].


\bibitem{BaBar-Kstarmumu}
  B.~Aubert {\it et al.}  [BABAR Collaboration],
  Phys.\ Rev.\  D {\bf 79}, 031102 (2009)
  [arXiv:0804.4412 [hep-ex]].

\bibitem{Aaltonen:2011ja}
  T.~Aaltonen {\it et al.} [ CDF Collaboration ],
  [arXiv:1108.0695 [hep-ex]].

\bibitem{Alok:2009tz}
  A.~K.~Alok, A.~Dighe, D.~Ghosh, D.~London, J.~Matias, M.~Nagashima, A.~Szynkman,
  JHEP {\bf 1002}, 053 (2010).
  [arXiv:0912.1382 [hep-ph]];  
  A.~K.~Alok, A.~Datta, A.~Dighe, M.~Duraisamy, D.~Ghosh and D.~London,
  JHEP {\bf 1111}, 121 (2011)
  [arXiv:1008.2367 [hep-ph]].


\bibitem{LHCbupdate2} Differential branching fraction and angular analysis of the $B \to K^{(*)} \mu^+ \mu^-$ decay, LHCb-CONF-2012-008.



\bibitem{delAguila:1985mk}
  F.~del Aguila and J.~Cortes,
  Phys.\ Lett.\  B {\bf 156}, 243 (1985).


\bibitem{Branco:1986my}
  G.~C.~Branco and L.~Lavoura,
  Nucl.\ Phys.\  B {\bf 278}, 738 (1986).

\bibitem{delAguila:1985ne}
  F.~del Aguila, M.~K.~Chase and J.~Cortes,
  Nucl.\ Phys.\  B {\bf 271}, 61 (1986).


\bibitem{Nir:1990yq}
  Y.~Nir and D.~J.~Silverman,
  Phys.\ Rev.\  D {\bf 42}, 1477 (1990).

\bibitem{Branco:1992wr}
  G.~C.~Branco, T.~Morozumi, P.~A.~Parada and M.~N.~Rebelo,
  Phys.\ Rev.\  D {\bf 48}, 1167 (1993).

\bibitem{Barger:1995dd}
  V.~D.~Barger, M.~S.~Berger and R.~J.~N.~Phillips,
  Phys.\ Rev.\  D {\bf 52}, 1663 (1995)
  [arXiv:hep-ph/9503204].

\bibitem{Silverman:1995rk}
  D.~Silverman,
  Int.\ J.\ Mod.\ Phys.\  A {\bf 11}, 2253 (1996)
  [arXiv:hep-ph/9504387].


\bibitem{Lavoura:1992qd}
  L.~Lavoura and J.~P.~Silva,
  Phys.\ Rev.\  D {\bf 47}, 1117 (1993).

\bibitem{Silverman:1991fi}
  D.~Silverman,
  Phys.\ Rev.\  D {\bf 45}, 1800 (1992).



\bibitem{Barenboim:1997pf}
  G.~Barenboim and F.~J.~Botella,
  Phys.\ Lett.\  B {\bf 433}, 385 (1998)
  [arXiv:hep-ph/9708209].

\bibitem{Barenboim:2000zz}
  G.~Barenboim, F.~J.~Botella and O.~Vives,
  Phys.\ Rev.\  D {\bf 64}, 015007 (2001)
  [arXiv:hep-ph/0012197].


\bibitem{Barenboim:2001}
  G.~Barenboim, F.~J.~Botella and O.~Vives,
  Nucl.\ Phys.\  B {\bf 613}, 285 (2001)
  [arXiv:hep-ph/0105306].


\bibitem{Chen:2010aq}
  C.~H.~Chen, C.~Q.~Geng and W.~Wang,
  JHEP {\bf 1011}, 089 (2010)
  [arXiv:1006.5216 [hep-ph]].

\bibitem{Alok:2010ij}
  A.~K.~Alok, S.~Baek and D.~London,
  JHEP {\bf 1107}, 111 (2011)
  [arXiv:1010.1333 [hep-ph]].

\bibitem{Botella:2012ju} 
  F.~J.~Botella, G.~C.~Branco and M.~Nebot,
  arXiv:1207.4440 [hep-ph].

\bibitem{Okada:2012gy} 
  Y.~Okada and L.~Panizzi,
  arXiv:1207.5607 [hep-ph].

\bibitem{minuit}
  F.~James and M.~Roos,
  Comput.\ Phys.\ Commun.\  {\bf 10}, 343 (1975).

\bibitem{Buras:1994dj}
  A.~J.~Buras and M.~Munz,
  Phys.\ Rev.\  D {\bf 52}, 186 (1995)
  [arXiv:hep-ph/9501281].


 \bibitem{Altmannshofer:2008dz}
  W.~Altmannshofer, P.~Ball, A.~Bharucha, A.~J.~Buras, D.~M.~Straub and M.~Wick,
  JHEP {\bf 0901}, 019 (2009)
  [arXiv:0811.1214 [hep-ph]].

\bibitem{Nir:1989rm}
  Y.~Nir,
  Phys.\ Lett.\  B {\bf 221}, 184 (1989).

 
\bibitem{buras1}
A.~J.~Buras, M.~Jamin and P.~H.~Weisz,
  Nucl.\ Phys.\  B {\bf 347}, 491 (1990).

\bibitem{LHCbupdate} 
  R.~Aaij {\it et al.}  [LHCb Collaboration],
  arXiv:1203.4493 [hep-ex].

\bibitem{CKMfitter}
Updates and numerical results of:
  The CKMfitter Group (J.~Charles {\it et al.}), 
  Eur.\ Phys.\ J.\  C {\bf 41}, 1 (2005)
  [arXiv:hep-ph/0406184],
available on the CKMfitter group web site:
\url{http://ckmfitter.in2p3.fr/}

\bibitem{Lenz:2012az} 
  A.~Lenz, U.~Nierste, J.~Charles, S.~Descotes-Genon, H.~Lacker, S.~Monteil, V.~Niess and S.~T'Jampens,
  arXiv:1203.0238 [hep-ph].


\bibitem{Aubert:2004it}
  B.~Aubert {\it et al.}  [BABAR Collaboration],
  Phys.\ Rev.\ Lett.\  {\bf 93}, 081802 (2004)
  [arXiv:hep-ex/0404006].

\bibitem{Iwasaki:2005sy}
  M.~Iwasaki {\it et al.}  [Belle Collaboration],
  Phys.\ Rev.\  D {\bf 72}, 092005 (2005)
  [arXiv:hep-ex/0503044].



\bibitem{Asner:2010qj}
  D.~Asner {\it et al.}  [Heavy Flavor Averaging Group],
  arXiv:1010.1589 [hep-ex].

\bibitem{Barate:2000rc}
  R.~Barate {\it et al.}  [ALEPH Collaboration],
  Eur.\ Phys.\ J.\  C {\bf 19}, 213 (2001)
  [arXiv:hep-ex/0010022].



\bibitem{pdg} K.~Nakamura {\it et al.}  [Particle Data Group],
  J.\ Phys.\ G {\bf 37}, 075021 (2010).


\bibitem{Blanke:2006ig}
  M.~Blanke, A.~J.~Buras, D.~Guadagnoli and C.~Tarantino,
  JHEP {\bf 0610}, 003 (2006)
  [arXiv:hep-ph/0604057].




\bibitem{Grossman:1995gt}
  Y.~Grossman, Z.~Ligeti and E.~Nardi,
  Nucl.\ Phys.\  B {\bf 465}, 369 (1996)
  [Erratum-ibid.\  B {\bf 480}, 753 (1996)]
  [arXiv:hep-ph/9510378].


\bibitem{Buras:1997fb}
  A.~J.~Buras and R.~Fleischer,
  Adv.\ Ser.\ Direct.\ High Energy Phys.\  {\bf 15}, 65 (1998)
  [arXiv:hep-ph/9704376].

\bibitem{Lenz:2006hd}
  A.~Lenz, U.~Nierste,
  JHEP {\bf 0706}, 072 (2007).
  [hep-ph/0612167].


 \bibitem{uli_lenz}
A.~Lenz, U.~Nierste,
    [arXiv:1102.4274 [hep-ph]].

\bibitem{Lenz:2012mb} 
  A.~Lenz,
  arXiv:1205.1444 [hep-ph].

\bibitem{Ali:1991is}
  A.~Ali, T.~Mannel and T.~Morozumi,
  Phys.\ Lett.\  B {\bf 273}, 505 (1991).

\end{thebibliography}
\end{document}